\RequirePackage{pdf14}

\documentclass[twocolumn]{autart}    

\usepackage{graphicx}      

\usepackage{amsmath}
\usepackage{amssymb}
\usepackage{commath} 
\usepackage{placeins}
\usepackage{multirow}
\usepackage{enumitem}  
\usepackage[normalem]{ulem}
\usepackage{xcolor}
\usepackage{cancel}
\usepackage{placeins}
\usepackage{caption}
\usepackage{ntheorem}
\usepackage{picins}
\graphicspath{{figures/}}

\newtheorem{theorem_th}{Theorem}
\newtheorem{lemma}{Lemma}
\newtheorem*{proof_th}{Proof}
\newcommand{\subparagraph}{}

\newcommand{\biographytwo}[3]{
\par\noindent 
\parbox[t]{\linewidth}{
\noindent\parpic{\includegraphics[width=.85in,keepaspectratio]{#1}}
\noindent {\scriptsize  {\bf #2}\
#3}}
}

\let\OLDthebibliography\thebibliography
\renewcommand\thebibliography[1]{
  \OLDthebibliography{#1}
  \setlength{\parskip}{0pt}
  \setlength{\itemsep}{0pt plus 0.3ex}
}

\begin{document}

\begin{frontmatter}

\title{Koopman Form of Nonlinear Systems with Inputs\thanksref{footnoteinfo}} 

\thanks[footnoteinfo]{This work has received funding from the European Research Council (ERC) under the European Union's Horizon 2020 research and innovation programme (grant agreement nr. 714663). This research has also been supported by the Eötvös Loránd Research Network (grant. number: SA-77/2021). The corresponding author is Lucian C. Iacob.}

\author[First]{Lucian C. Iacob}\ead{l.c.iacob@tue.nl},    
\author[First,Second]{Roland T\'oth}, 
\author[First]{Maarten Schoukens}

\address[First]{Control Systems Group, Eindhoven University of Technology, Eindhoven, The Netherlands}  
\address[Second]{Systems and Control Lab, Institute for Computer Science and Control, Budapest, Hungary}             


\begin{abstract}
\vskip -3mm \noindent                         
The Koopman framework proposes a linear representation of finite-dimensional nonlinear systems through a generally infinite-dimensional globally linear embedding. Originally, the Koopman formalism has been derived for autonomous systems. In applications for systems with inputs, generally a \emph{linear time invariant} (LTI) form of the Koopman model is assumed, as it facilitates the use of control techniques such as linear quadratic regulation and model predictive control.  However, it can be easily shown that this assumption is insufficient to capture the dynamics of the underlying nonlinear system. Proper theoretical extension for actuated continuous-time systems 
with a linear or a control-affine input has been worked out only recently, however extensions to discrete-time systems and general continuous-time systems  have not been developed yet.
In the present paper, we systematically investigate and analytically derive lifted forms under inputs for a rather wide class of nonlinear systems in both continuous and discrete time. 
We prove that the resulting lifted representations give Koopman
models where the state transition is linear, but the input matrix becomes state-dependent (state
and input-dependent in the discrete-time case), giving rise to a specially structured \emph{linear parameter-varying} (LPV) description of the underlying system. We also provide error bounds on how much the dependency of the input matrix contributes to the resulting representation and how well the system behaviour can be approximated by an LTI Koopman representation. The introduced theoretical insight greatly helps for performing proper model structure selection in system identification with Koopman models as well as making a proper choice for LTI or LPV techniques for the control of nonlinear systems through the Koopman approach. \vspace{-4mm}
\end{abstract}

\end{frontmatter}

\section{Introduction}
\smallskip
Nowadays, most dynamic systems exhibit nonlinear behaviour that needs to be handled to meet the ever increasing performance requirements. Nonlinear control techniques, e.g., backstepping, control Lyapunov functions \cite{Isidori:95,Khalil:02}, are complex and generally only provide stability guarantees. In contrast, the control methods developed for linear systems offer strong guarantees on performance and easy to use design principles. However, using linearized models allows only limited performance, as they are only valid locally. For this reason, in recent years, there has been a significant research interest on embedding nonlinear systems into linear models. A possible way to achieve this is through the Koopman framework \cite{Koopman:31} that proposes a trade-off in complexity by representing a nonlinear system through a generally infinite-dimensional, but linear description. By applying a nonlinear state transformation through so-called observable functions, the states are projected into a higher dimensional space where their dynamic relation can be expressed as a linear mapping. While the framework is well worked out for autonomous systems, treating systems with inputs is not understood well. 
\par In the \emph{continuous-time} (CT) case, for input-linear or control-affine nonlinear systems, applying the chain rule of differentiation leads to not entirely linear Koopman models \cite{Kaiser:20}. More specifically, using state-dependent observables results in a constant state-transition matrix, while the input matrix becomes state-varying. Hence, the Koopman form can be interpreted as a \emph{linear parameter-varying} (LPV) model \cite{Mohammadpour:12}.  Furthermore, if the resulting input matrix function is in the span of the observables, the resulting Koopman model associated with a nonlinear control-affine system can be written in a bilinear form \cite{Goswami:17,Huang:18}. How to systematically obtain Koopman forms suitable for control of general nonlinear systems with input is still an open question even for 
CT systems. For example, the authors of \cite{Klus:20} consider a finite set of discrete input values and derive a family of Koopman generators which are defined only for these specific constant inputs. This results in a switched linear system and the control approach aims to optimize the switching sequence. Alternatively, one could consider the Koopman operator acting on an extended state-input space \cite{Kaiser:20}. However, the resulting lifted model is autonomous in the extended state and the construction is difficult to be used for control purposes. As such, the current available methods to derive an associated Koopman form of nonlinear systems with input in a general form offer limited control possibilities.
\par In \emph{discrete time} (DT), the derivation of Koopman models is not as straightforward even for linear input and control-affine nonlinear systems due to the lack of chain rule for the difference operator. However, for system identification and embedded control purposes, the 
DT models are predominantly used. In the literature, there are several different methods that treat 
DT nonlinear systems with inputs. For example, the authors of \cite{Williams:16} suggest to identify an autonomous LPV Koopman model with the input representing the scheduling variable. A more commonly used approach is detailed in the work of \cite{Proctor:18}, which uses a dictionary of state, input and mixed-dependent observables. While the representation is autonomous, it is common to restrict the output space of the Koopman operator to observables only dependent on the state. This idea has been used mostly in identification-related works, e.g. \cite{Bonnert:20,Liu:18}, and works that investigate the relation of system theoretic properties between the Koopman form and the original nonlinear representation, such as \cite{Yeung:18}. Furthermore, the input might also be projected through the nonlinear lifting, which complicates the control problem. As an alternative to observable-based lifting, several works, e.g., \cite{Kaiser:20,Korda:20}, use the spectral properties of the Koopman operator and lift the autonomous dynamics using eingenfunction coordinates. However, the problem of deriving Koopman forms with direct inputs still remains unresolved.
\par  Given its simplicity, a \emph{linear time invariant} (LTI) Koopman model with nonlifted input is generally used in practice, especially for control methods such as \emph{linear quadratic regulation} (LQR) and \emph{model predictive control} (MPC), \cite{Korda:18,Mamakoukas:20,Ping:21}. However, there is no discussion on the validity of these models or on the approximation error that is introduced. Furthermore, as we show in this paper, the involved approximation error can be substantial. There also exist works that use models of different complexities, e.g. bilinear \cite{Zinage:23}, control-affine or nonlinear input function \cite{Shi:22}, or employ reinforcement learning techniques together with an autonomous Koopman model with observables dependent on the state and policy parameters \cite{Zanini:21}. However, while the control applications show promising results, it is clear that a theory to formulate exact Koopman models under inputs that are useful for control purposes is lacking.
\par Based on these, the present paper investigates the analytic derivation of Koopman forms of general nonlinear systems with inputs both in 
CT and DT. 
The contributions can be summarized as follows.
\begin{itemize}
\item[C1] A systematic factorization method 
is devised to obtain, for a wide class of  nonlinear systems, an exact continuous-time Koopman model suitable for control purposes.
\item[C2] A method to analytically compute an exact Koopman form of a wide class of discrete-time nonlinear systems with inputs is developed.
\item[C3] Providing an interpretation of the resulting Koopman forms as LPV models both in continuous and discrete time.
\item[C4] A 2-norm based magnitude bound of the state response error between the derived exact Koopman model and its LTI approximations is devised to give characterization of the expected model uncertainty.
\end{itemize} 
\par The paper is structured as follows. Section \ref{sec:CT} gives an introduction to Koopman forms in 
CT and presents our Contributions C1 and C3 in terms of a general approach for obtaining Koopman representations under inputs together with their LPV interpretation. In Section \ref{sec:dt_koop}, we further extend our results for the Koopman framework to 
DT systems, which is the main contribution (see C2 and C3) of our work. In Section \ref{sec:section_error}, we develop an error bound to characterize how well the system behaviour can be captured using an LTI Koopman model, constituting Contribution C4. Next, in Section \ref{sec:examples}, we apply our proposed approach to nonlinear systems both in 
CT and DT
to demonstrate 
the utilization potential of the developed theory. 
Finally, in Section \ref{sec:conclusion}, conclusions on the presented results are drawn.
\section{Embedding of continuous-time systems}\label{sec:CT}
\smallskip
\vspace{-.1cm}
The present section briefly details the embedding of autonomous 
CT nonlinear systems into Koopman forms, followed by the extension of the embedding under inputs. Additionally, we propose a factorization method to describe the resulting Koopman models in an LPV form which is useful for control.
\subsection{Koopman representation of autonomous systems}

Consider an 
autonomous CT nonlinear time-invariant system: \vspace{-1.5mm}
\begin{equation}\label{eq:aut_ct}
\dot{x}_t=f_\mathrm{c}(x_t), \quad x_0\in \mathbb{X}\subseteq \mathbb{R}^{n_\mathrm{x}}, \vspace{-0mm}
\end{equation}
where $f_\mathrm{c}:\mathbb{R}^{n_\mathrm{x}}\rightarrow\mathbb{R}^{n_\mathrm{x}}$ is Lipschitz continuous, $x_t:=x(t)$ is the state variable and $t\in\mathbb{R}$ denotes the continuous time. The solution $x_t$ of \eqref{eq:aut_ct} at time $t$ starting from an initial condition $x_0$ can be described by the induced flow:\vspace{-.3cm}
\begin{equation}\label{eq:flow_ct}
x_t= F_\mathrm{c}(t,x_0)=x_0 + \int^{t}_{0}f_\mathrm{c}(x_\tau)\dif \tau.
\vspace{-.3cm}
\end{equation}
The set $\mathbb{X}$ is considered to be compact and forward invariant under $F_\mathrm{c}(t,\cdot)$, i.e. $F_\mathrm{c}(t,\mathbb{X})\subseteq \mathbb{X}$, $\forall t\geq 0$ (assuming a weak form of stability). The Koopman family of operators $\left\lbrace\mathcal{K}^t\right\rbrace_{t\geq 0}$ associated with $F_\mathrm{c}(t,\cdot)$ for a given $\mathcal{F} \subseteq  \mathcal{C}^1$ (continuously differentiable) Banach function space is defined as the composition:\vspace{-.2cm}
\begin{equation}\label{eq:koop_ct_def}
\mathcal{K}^t \phi(x_0)=\phi\circ F_\mathrm{c}(t,x_0),\quad \forall \phi\in\mathcal{F},
\vspace{-.2cm}
\end{equation}
where $\phi:\mathbb{X}\rightarrow \mathbb{R}$ denotes scalar observable functions in $\mathcal{F}$.
As shown in \cite{Mauroy:20}, the family $\{\mathcal{K}^t\}_{t\geq 0}$ is a (one-parameter) semigroup and characterizes the linear embedding of \eqref{eq:aut_ct} in terms of the observables $\phi\in\mathcal{F}$. Furthermore, as $\mathbb{X}$ is a compact forward-invariant set and the flow $F_\mathrm{c}$ is uniformly Lipschitz continuous w.r.t. $t$, the Koopman semigroup $\{ \mathcal{K}^t\}_{t\geq 0}$ is strongly continuous on $\mathcal{F}$ \cite{Mauroy:20}. Hence, the infinitesimal generator  $\mathcal{L}:\mathcal{D}_\mathcal{L}\rightarrow \mathcal{F}$ of the Koopman semigroup of operators is defined as  \vspace{-.2cm}
\begin{equation}\label{eq:lim_generator_ct}
\mathcal{L}\phi(x_0)=\lim_{t\downarrow 0}\frac{\mathcal{K}^t\phi(x_0)-\phi(x_0)}{t},\;\; \forall\phi\in\mathcal{D}_\mathcal{L},\vspace{-.2cm}
\end{equation} 
see \cite{Lasota:94,Mauroy:20}, where the domain $\mathcal{D}_\mathcal{L}\subseteq \mathcal{F}$ is a dense set in $\mathcal{F}$ and the limit \eqref{eq:lim_generator_ct} exists in the strong sense \cite{Lasota:94}:\vspace{-.2cm}
\begin{equation}
\lim_{t\downarrow 0} \left\Vert \mathcal{L}\phi(x_0) - \frac{\mathcal{K}^t \phi(x_0) - \phi(x_0)}{t} \right\Vert = 0,\vspace{-.2cm}
\end{equation}
with $\|\cdot\|$ being the norm associated with $\mathcal{F}$ \cite{Mauroy:20}. In 
CT, the notion of the infinitesimal generator is important, as it is used to describe the dynamics in the lifted space of observables. As shown in \cite{Engel:00}, if the previous assumptions hold true, then the generator $\mathcal{L}$ is linear. Let $z_\phi(t)=\mathcal{K}^t\phi(x_0)$. Then, $z_\phi(t)$ is the solution of the equation: \vspace{-3mm}
\begin{equation}\label{eq:ivp_ct}
\dot{z}_\phi=\mathcal{L}z_\phi,\vspace{-2mm}
\end{equation}
with initial condition $z_\phi(0)=\phi(x_0)$ \cite{Bakker:19}. Solving the initial value problem for \eqref{eq:ivp_ct} results in  
$\mathcal{K}^t =e^{\mathcal{L}t}$. 
As described in \cite{Bakker:19,Bevanda:21}, the application of 
$\mathcal{L}$ on 
$\phi(x_t)$ gives:\vspace{-.3cm}
\begin{equation}\label{eq:obs_ct_generator}
\dot{\phi}=\frac{\partial \phi}{\partial x}f_\mathrm{c}=\mathcal{L}\phi, \vspace{-.2cm}
\end{equation}
which defines a linear, but infinite-dimensional representation of the underlying system. In this paper, we consider that there exists a finite-dimensional Koopman invariant subspace $\mathcal{F}_{n_\mathrm{f}}\subseteq \mathcal{D}_\mathcal{L}$, i.e., $\mathcal{L}:\mathcal{F}_{n_\mathrm{f}}\rightarrow\mathcal{F}_{n_\mathrm{f}}$. As shown in \cite{Bakker:19}, if $\mathcal{F}_{n_\mathrm{f}}$ is invariant under the Koopman generator $\mathcal{L}$, then, due to the linearity of $\mathcal{L}$, $\mathcal{L}\phi$ is a linear combination of the elements of $\mathcal{F}_{n_\mathrm{f}}$. Consider $\Phi^\top=[\ \phi_1\ \cdots\ \phi_{n_\mathrm{f}}\ ]$ to be a basis of $\mathcal{F}_{n_\mathrm{f}}$. Following the derivations in \cite{Mauroy:20}, 
the effect of the infinitesimal generator on a component $\phi_j$ can be described as:\vspace{-.4cm}
\begin{equation} 
\dot{\phi}_j=\mathcal{L} \phi_j=\sum^{n_\mathrm{f}}_{i=1}L_{i,j}\phi_i,\vspace{-.3cm}
\end{equation}
where $L$ is a matrix representation of the Koopman generator and the $j^{\text{th}}$ column of $L$ contains the coordinates of $\mathcal{L} \phi_j$ in the basis $\Phi$. By introducing $A=L^\top \in \mathbb{R}^{n_\mathrm{f}\times n_\mathrm{f}}$, the lifted form of the dynamics of \eqref{eq:aut_ct} can be written as:\vspace{-.2cm}
\begin{equation}\label{eq:ct_koopman_aut}
\dot{\Phi}(x_t)=A\Phi(x_t).\vspace{-.2cm}
\end{equation}
Based on \eqref{eq:obs_ct_generator}, the following relation also holds true:\vspace{-.3cm}
\begin{equation}\label{eq:ct_koop_aut2}
\dot{\Phi}(x_t)=\frac{\partial \Phi}{\partial x}(x_t)f_\mathrm{c}(x_t),\vspace{-.3cm}
\end{equation}
where $\frac{\partial \Phi}{\partial x}$ is the Jacobian of $\Phi$. Hence, the generally used condition in the literature to have a finite-dimensional Koopman embedding (i.e. lifting) for \eqref{eq:aut_ct} is to find a set of observables $\Phi$ for which:\vspace{-2mm}
\begin{equation}\label{eq:span_condition_ct}
\frac{\partial \Phi}{\partial x}f_\mathrm{c}\in \text{span} \left\lbrace\Phi\right\rbrace. \vspace{-.2cm}
\end{equation}
To recover the original states, the existence of an inverse transformation $\Phi^\dagger(\Phi(x_t))=x_t$ is often assumed. In practice, this is 
achieved by either including the states as part of the observables or by requiring that they are in the span of the observables.
\par Finally, to explicitly give the LTI dynamics implied by the  lifted form, introduce $z_t=\Phi(x_t)$, which gives the Koopman representation of \eqref{eq:aut_ct} as:\vspace{-.2cm}
\begin{equation}
\dot{z}_t=Az_t, \qquad \text{with } z_0=\Phi(x_0).\vspace{-.2cm}
\end{equation}
\subsection{Koopman representation under inputs}
\smallskip
The following section treats the derivation of the Koopman representation for systems with inputs. The approach is based on a sequential method that uses state-dependent observables, as done in \cite{Kaiser:20,Surana:16}, for control-affine systems and in \cite{Surana:16} for nonlinear systems in general form. As a new result, we describe a factorization method for general nonlinear systems to obtain a model useful for control. Furthermore, we interpret the resulting Koopman representations as LPV models.
\subsubsection{General nonlinear systems}
\smallskip
Consider the CT nonlinear time-invariant system: \vspace{-2mm}
\begin{equation}\label{eq:ct_nl_general}
\dot{x}_t=f_\mathrm{d}(x_t,u_t) \vspace{-2mm}
\end{equation}
with $x_0 \in \mathbb{X}$, $u_t\in\mathbb{U}\subseteq\mathbb{R}^{n_\mathrm{u}}$, and $f_\mathrm{d}:\mathbb{R}^{n_\mathrm{x}}\times\mathbb{U}\rightarrow \mathbb{R}^{n_\mathrm{x}}$ is Lipschitz continuous. It is assumed that $\mathbb{U}$ is given such that $\mathbb{X}$ is compact and forward  invariant under the induced flow. 
To avoid ending up with a Koopman model of \eqref{eq:ct_nl_general} without explicit external inputs that is difficult to control, we use a construction based on only state-dependent observables. First, we decompose the function $f_\mathrm{d}(x_t,u_t)$ into the sum between the contributions of the autonomous and input-related dynamics:\vspace{-.2cm}
\begin{equation}\label{eq:ct_general_sum}
f_\mathrm{d}(x_t,u_t)=\underbrace{f_{\mathrm{d}}(x_t,0)}_{f_\mathrm{c}(x_t)}  + \underbrace{f_\mathrm{d}(x_t,u_t) - f_{\mathrm{d}}(x_t,0)}_{g_\mathrm{c}(x_t,u_t)},  
\vspace{-.2cm}
\end{equation}
 where $g_\mathrm{c}(x_t,0)=0$. Note that decomposition \eqref{eq:ct_general_sum}, which has also been described in \cite{Surana:16}, is trivial and always exists for any $f_\mathrm{d}$.
Next, we give the analytically derived Koopman representation associated with \eqref{eq:ct_nl_general}. 
\vskip 1mm
\begin{theorem_th}\label{thm:theorem_ct}
Given a nolinear CT system in the general form \eqref{eq:ct_nl_general}, where $f_\mathrm{d}$ is written as \eqref{eq:ct_general_sum}, 
with 
the observables $\Phi:\mathbb{X}\rightarrow\mathbb{R}^{n_\mathrm{f}}$ 
in $\mathcal{C}^1$ such that \eqref{eq:span_condition_ct} holds for $f_{\mathrm{c}}(x_t)$, then there exists an exact finite-dimensional lifted form \vspace{-.3cm}
\begin{subequations}
\begin{equation}\label{eq:koop_ct_general_model}
\dot{\Phi}(x_t)=A\Phi(x_t)+\mathcal{B}(x_t,u_t),\vspace{-.2cm}
\end{equation}
with $A\in\mathbb{R}^{n_\mathrm{f}\times n_\mathrm{f}}$ and $\mathcal{B}: 
\mathbb{X}\times\mathbb{U}\rightarrow \mathbb{R}^{n_{\mathrm{f}}}$ defined as:\vspace{-.3cm}
\begin{equation}\label{eq:ct_B_tilde_general}
\mathcal{B}(x_t,u_t)=\frac{\partial \Phi}{\partial x}(x_t)g_{\mathrm{c}}(x_t,u_t).\vspace{-.2cm}
\end{equation}
\end{subequations}
\end{theorem_th}
\begin{proof_th}
Based on \eqref{eq:span_condition_ct}, lifting the autonomous dynamics gives the finite-dimensional lifted representation form:\vspace{-.2cm}
\begin{equation}\label{eq:lifting_phi}
\dot{\Phi}(x_t)=\frac{\partial \Phi}{\partial x}(x_t)f_\mathrm{c}(x_t) =A\Phi(x_t).\vspace{-.3cm}
\end{equation}
Next, for $u_t\neq 0$, the dynamics are described by \eqref{eq:ct_general_sum}. Taking the time derivative of 
$\Phi$, 
the chain rule gives: 
\vspace{-.2cm}
\begin{align}\label{eq:ct_general_case} 
\dot{\Phi}(x_t)&=\frac{\partial \Phi}{\partial x}(x_t)f_\mathrm{d}(x_t,u_t) \notag \\
&=\frac{\partial \Phi}{\partial x}(x_t)f_\mathrm{c}(x_t) +\frac{\partial \Phi}{\partial x}(x_t)g_\mathrm{c}(x_t,u_t) \notag \\
&=A\Phi(x_t)+\mathcal{B}(x_t,u_t) 
\end{align}
\vskip -3mm
with $\mathcal{B}$ given by \eqref{eq:ct_B_tilde_general}. \hfill $\blacksquare$ 
\end{proof_th}
Eq. \eqref{eq:koop_ct_general_model} represents an exact lifted form of  \eqref{eq:ct_nl_general}, consistent with the lifting in \eqref{eq:lifting_phi} under $\Phi$. Note that the input enters through the nonlinear function $\mathcal{B}(x_t,u_t)$, which limits the application of \eqref{eq:koop_ct_general_model} for control purposes. To address this, we can recast the representation into a so-called LPV form by the help of the following Lemma:
\vskip 1mm
\begin{lemma}\label{lma_fact}
Let $\mathcal{B}: 
\mathbb{X} \times \mathbb{U}\rightarrow\mathbb{R}^{n_\mathrm{f}}$ be continuously differentiable in $u_t$, continuous in $x_t$ and satisfying $\mathcal{B}(x_t,0)=0$, and let $\mathbb{U}$ be a convex set containing the origin\footnote{For most real-world systems, convexity of $\mathbb{U}$ is not a strong assumption due to limited input range of actuators or operational limits of the system. If $0\not \in \mathbb{U}$, then, by re-centering $u$ and modifying $f_\mathrm{d}$ accordingly, the condition can be satisfied.}. Then,\vspace{-.3cm}
\begin{equation}\label{eq:lma_fact_eq}
B(x_t,u_t)=\int^1_0 \frac{\partial \mathcal{B}}{\partial u}(x_t,\lambda u_t)\dif \lambda,\vspace{-.3cm}
\end{equation}
provides a factorization of $\mathcal{B}$ such that $\mathcal{B}(x_t,u_t)=B(x_t,u_t)u_t$ for any $(x_t,u_t)\in(\mathbb{X},\mathbb{U})$. 
\end{lemma}
\vskip 1mm
\begin{proof_th}
See Appendix \ref{apx:proof_lemma}. 
\end{proof_th}
\par The resulting lifted representation for continuous-time nonlinear systems in general form is: \vspace{-.2cm}
\begin{equation}\label{eq:koop_general_ct_factorized}
\dot{\Phi}(x_t)=A\Phi(x_t)+B(x_t,u_t)u_t. \vspace{-.2cm}
\end{equation}
To handle the possibly nonlinear dependency of the input matrix $B(x_t,u_t)$ on the state $x_t$ and input $u_t$, we can express \eqref{eq:koop_general_ct_factorized} in an LPV form. Let $z_t=\Phi(x_t)$ and introduce a so called scheduling map, $p_t=\mu(z_t,u_t)$, such that $B_\mathrm{z} \circ \mu =B$ and $B_\mathrm{z}$ belongs to a predefined function class, such as affine, polynomial, rational. Then, we can introduce an LPV form of the Koopman model of \eqref{eq:ct_nl_general} as\vspace{-.2cm}
\begin{equation}\label{eq:lpv_koop_ct}
\dot{z}_t=Az_t + B_\mathrm{z}(p_t)u_t,\vspace{-.2cm}
\end{equation}
with $z_0=\Phi(x_0)$. LPV forms such as \eqref{eq:lpv_koop_ct} are well suited to address nonlinear control problems by the use of powerful convex optimization based analysis of stability and performance together with synthesis tools to obtain controllers, observers etc. with performance guarantees, see \cite{BriatBook,Mohammadpour:12,Toth:12}. To arrive to an LPV form, the choice of $\mu$
is often driven by the utilization of the resulting model. Choosing a $\mu$ such that $B(x_t,u_t)$ is transformed to a $B_\mathrm{z}$ that is an affine function of $p_t$ is highly advantageous as it enables the use of the most simplest and often computationally efficient polytopic analysis and synthesis tools. However, such a choice for $\mu$ can result in a high dimension of $p_t$, which increases the conservativeness affecting the analysis and may render the synthesis unfeasible. Alternatively, a much smaller dimension for $p_t$ can be achieved by considering a function $\mu$ such that $B_\mathrm{z}$ becomes polynomial or even rational. For such dependency classes, there is also an extensive array of advanced analysis and synthesis tools available based on linear fractional representations, $\mu$-analysis, full-block multipliers, and integral quadratic constraints \cite{BriatBook,Mohammadpour:12}. 
\subsubsection{Control-affine or linear input cases}
\smallskip
In case \eqref{eq:ct_nl_general} is in a control-affine form:\vspace{-.15cm}
\begin{equation}\label{eq:ct_nl_ca}
\dot{x}_t=f_\mathrm{c}(x_t)+g_\mathrm{c}(x_t)u_t,\vspace{-.15cm}
\end{equation}
with $g_\mathrm{c}:\mathbb{R}^{n_\mathrm{x}}\rightarrow\mathbb{R}^{n_\mathrm{x}\times n_\mathrm{u}}$ and $u_t\in\mathbb{U}\subseteq\mathbb{R}^{n_\mathrm{u}}$, Theorem \ref{thm:theorem_ct} can be applied to obtain the lifted representation:\vspace{-.25cm}
\begin{equation}\label{eq:koop_ca_ct}
\dot{\Phi}(x_t)=A\Phi(x_t)+B(x_t)u_t,\vspace{-.25cm}
\end{equation}
where $B(x_t)u_t = \mathcal{B}(x_t,u_t)$ and \vspace{-.25cm}
\begin{equation}
B(x_t) = \frac{\partial \Phi}{\partial x}(x_t)g_\mathrm{c}(x_t).\vspace{-.25cm}
\end{equation}
Again, the lifted representation \eqref{eq:koop_ca_ct} can be expressed in an LPV form, but with only a state-dependent scheduling variable, i.e. $p_t=\mu(z_t)$.
Furthermore, nonlinear systems with linear input ($g_\mathrm{c}(x_t)=b\in\mathbb{R}^{n_\mathrm{x}\times n_\mathrm{u}}$ is a constant matrix) 
are a particular case of \eqref{eq:ct_nl_ca}, hence the associated lifted form is also described by \eqref{eq:koop_ca_ct} and the Koopman representation \eqref{eq:lpv_koop_ct}. Note that the input matrix $B$ is still state-dependent due to the multiplication with the Jacobian function, as $B(x_t)=\frac{\partial \Phi}{\partial x}(x_t)b$.
\par As discussed in \cite{Goswami:17,Huang:18,Schulze:22}, if $\frac{\partial \Phi}{\partial x}g_{\mathrm{c}_i} \in \text{span} \left\lbrace \Phi \right\rbrace$ with $g_{c_i}$ being the $i^{\text{th}}$ column of $g_\mathrm{c}$, then there is a 
$B_i\in\mathbb{R}^{n_\mathrm{f}\times n_\mathrm{f}}$ satisfying 
$\frac{\partial \Phi}{\partial x}g_{\mathrm{c}_i}=B_i\Phi$, 
giving the bilinear  lifted form:\vspace{-.2cm}
\begin{equation}\label{eq:bilinear_form_ct}
\dot{\Phi}(x_t)=A\Phi(x_t)+\sum^{n_{\mathrm{u}}}_{i=1}B_i\Phi(x_t)u_{t,i}.\vspace{-.15cm}
\end{equation}
with $u_{t,i}$ being the $i^{\text{th}}$ element of $u_t$. Equivalently, let $z_t=\Phi(x_t)$, then a bilinear Koopman model of \eqref{eq:ct_nl_ca} is:\vspace{-.4cm}
\begin{equation}\label{eq:bilinear_koopman_v1}
\dot{z}_t = Az_t + \sum^{n_{\mathrm{f}}}_{j=1}z_{t,j}\tilde{B}_j u,\vspace{-.4cm}
\end{equation}
where $z_0 = \Phi(x_0)$, $z_{t,j}$ is the $j^{\text{th}}$ element of $z_t$, and $\tilde{B}_j=[\ B_{1,j}\ \cdots \ B_{n_{\mathrm{u}},j}\ ]$ where $B_{i,j}\in\mathbb{R}^{n_{\mathrm{f}}}$ is the $j^{\text{th}}$ column of $B_i$.
\vspace{-.1cm}
\section{Embedding of discrete-time systems}\label{sec:dt_koop}
\smallskip
\vspace{-.1cm}
In this section, we extend the previously described results for 
DT systems, which is our main contribution.
\subsection{Koopman representation of autonomous systems}
\smallskip
\vspace{-.1cm}
Consider the DT autonomous time-invariant nonlinear system:\vspace{-.1cm}
\begin{equation}\label{eq:aut_dt}
x_{k+1} = f(x_k),
\end{equation}
with initial condition $x_0\in \mathbb{X}\subseteq\mathbb{R}^{n_\mathrm{x}}$, nonlinear state transition map $f:\mathbb{R}^{n_\mathrm{x}}\rightarrow\mathbb{R}^{n_\mathrm{x}}$ and $k \in \mathbb{Z}$ the discrete time. It is assumed that $\mathbb{X}$ is compact and forward invariant under $f(\cdot)$, i.e. $f(\mathbb{X})\subseteq \mathbb{X}$. For DT systems, as expressed in \cite{Mauroy:16}, the DT Koopman operator is generally 'fixed' to the sampling interval, i.e. $\mathcal{K}^{t}$, for $t=T_\mathrm{s}$, to describe the evolution of the observables between each time step. We drop the superscript to ease readability. The Koopman operator $\mathcal{K}:\mathcal{F}\rightarrow\mathcal{F}$ associated with the nonlinear map $f$ is defined through the composition:\vspace{-.2cm}
\begin{equation}\label{eq:koop_dt_comp}
\mathcal{K}\phi=\phi\circ f, \qquad \forall\phi\in\mathcal{F},\vspace{-.2cm}
\end{equation}
where $\mathcal{F}$ is a Banach function space of observables \linebreak  $\phi:\mathbb{X}\rightarrow \mathbb{R}$. Given an arbitrary state $x_k\in\mathbb{X}$, \eqref{eq:koop_dt_comp} is equivalent to the following relation:\vspace{-.2cm}
\begin{equation}
\mathcal{K}\phi(x_k)=\phi\circ f(x_k)=\phi(x_{k+1}).\vspace{-.2cm}
\end{equation} 
We assume there exists a finite-dimensional Koopman invariant subspace $\mathcal{F}_{n_\mathrm{f}}\subseteq \mathcal{F}$, i.e., $\mathcal{K}:\mathcal{F}_{n_\mathrm{f}}\rightarrow \mathcal{F}_{n_\mathrm{f}}$. As $\mathcal{K}$ is a linear operator \cite{Mauroy:20}, $\mathcal{K}\phi$ can be expressed as a linear combination of the elements of $\mathcal{F}_{n_\mathrm{f}}$. Let $\Phi^\top\!=[\ \phi_1\ \cdots\ \phi_{n_\mathrm{f}}\ ]$ be a basis of $\mathcal{F}_{n_\mathrm{f}}$. As detailed in \cite{Mauroy:20}, the effect of the Koopman operator on 
$\phi_j$ is expressed as:\vspace{-.3cm}
\begin{equation}
\mathcal{K}\phi_j=\sum^{n_\mathrm{f}}_{i=1}K_{i,j}\phi_i,\vspace{-.3cm}
\end{equation}
where $K$ is the matrix representation of the Koopman operator and the $j^{\text{th}}$ column of $K$ contains the coordinates of $\mathcal{K}\phi_j$ in the basis $\Phi$. By taking $A=K^\top$, a finite-dimensional representation of \eqref{eq:aut_dt} in the lifted space is \vspace{-.2cm}
\begin{equation}\label{eq:dt_koop_aut_1}
\Phi(x_{k+1})=A\Phi(x_k). \vspace{-.2cm}
\end{equation}
Based on \eqref{eq:aut_dt}, we can substitute the LHS of \eqref{eq:dt_koop_aut_1} in terms of $\Phi(x_{k+1}) = \Phi(f(x_k))$ giving:\vspace{-.2cm}
\begin{equation}\label{eq:dt_koop_aut}
\Phi\circ f(x_k)=A\Phi(x_k).\vspace{-.2cm}
\end{equation}
Based on \eqref{eq:dt_koop_aut}, the existence condition of a Koopman invariant subspace for a given choice of observables is
\vspace{-.2cm}
\begin{equation}\label{eq:dt_span_condition}
\Phi\circ f \in \text{span}\left\lbrace\Phi\right\rbrace.\vspace{-.2cm}
\end{equation}
Similar to the CT case, we can ensure the existence of an inverse transformation $\Phi^\dagger(\Phi(x_k))=x_k$ by requiring the identity function $x=\mathrm{id}(x)$ to also be in the span of $\Phi$, i.e., $\mathrm{id}\in\mathrm{span}\{\Phi\}$. 
Given $z_k = \Phi(x_k)$, the LTI Koopman model associated with \eqref{eq:aut_dt} is:\vspace{-.2cm}
\begin{equation}
z_{k+1}=Az_k, \qquad \text{with } z_0=\Phi(x_0).\vspace{-.2cm}
\end{equation}
\subsection{Koopman representation under inputs}
\smallskip
Next, we derive a Koopman representation for DT systems with input. Compared to the CT case, the chain rule can no longer be applied to derive the lifted representation. Hence, we propose a method based on the FTC, to analytically derive the DT lifted form. 
\subsubsection{General nonlinear systems}
\smallskip
Consider a DT nonlinear system in the general form:\vspace{-.2cm}
\begin{equation}\label{eq:dt_nl_general}
x_{k+1}=f_{\mathrm{d}}(x_k,u_k),\vspace{-.2cm}
\end{equation}
with $u_k\in\mathbb{U}\subseteq \mathbb{R}^{n_\mathrm{u}}$ being the input, $x_0\in\mathbb{X}$ and $f_{\mathrm{d}}:\mathbb{R}^{n_\mathrm{x}} \times \mathbb{U}\rightarrow \mathbb{R}^{n_\mathrm{x}}$. It is assumed that $\mathbb{U}$ is given such that $\mathbb{X}$ is compact and forward invariant under $f$.
As in the CT case, we propose an approach that uses only state-dependent observables to analytically derive the Koopman form. 
Similar to the 
CT case, we decompose $f_{\mathrm{d}}(x_k,u_k)$ to autonomous and input-driven dynamics:\vspace{-.2cm}
\begin{equation} \label{DT:sep}
f_{\mathrm{d}}(x_k,u_k)=f(x_k)+g(x_k,u_k).\vspace{-.2cm}
\end{equation} 
The first step is to lift the autonomous dynamics by considering zero input, i.e., $u_k=0$:\vspace{-.2cm}
\begin{equation}
x_{k+1}=f_{\mathrm{d}}(x_k,0)=f(x_k),\vspace{-.2cm}
\end{equation}
and $g(x_k,0)=0$. Applying the function $\Phi$, the lifted dynamics of the autonomous part can be written as follows, assuming \eqref{eq:dt_span_condition} is satisfied:\vspace{-.2cm}
\begin{equation}\label{eq:dt_koop_aut_gen}
\Phi(x_{k+1})=A\Phi(x_k)=\Phi(f(x_k)).\vspace{-.2cm}
\end{equation}
Next, by considering the full dynamics of \eqref{eq:dt_nl_general}, we apply the lifting $\Phi$:\vspace{-.2cm}
\begin{equation}
\Phi(x_{k+1})= \Phi\bigl(f(x_k) + g(x_k,u_k)\bigr).\vspace{-.2cm}
\end{equation}
In contrast to the CT case, the absence of a chain rule under the shift operator makes it hard to 
directly separate the autonomous and input-driven contributions. To solve this problem, we employ the FTC to derive analytically an exact Koopman representation. \smallskip
\begin{theorem_th}\label{thm:theorem_dt}
Given a nonlinear DT system in the general form  \eqref{eq:dt_nl_general}, where $f_{\mathrm{d}}$ is written as \eqref{DT:sep}, together with 
observables $\Phi:\mathbb{X}\rightarrow\mathbb{R}^{n_\mathrm{f}}$ 
in $\mathcal{C}^1$ with $\mathbb{X}$ convex,\footnote{To satisfy the  convexity requirement, 
one can always construct a convex forward invariant set in $\mathbb{X}$. 
} such that $\Phi(f(\cdot))\in \text{span}\left\lbrace\Phi\right\rbrace$\footnote{This condition is taken to simplify the mathematical setting in terms of finite dimensional operators. However, the implication 
of the results in this paper are also valid for the Koopman form under inputs in an approximate sense (when $\Phi(f(\cdot))\notin \text{span}\left\lbrace\Phi\right\rbrace$). 
}, then there exists an exact finite-dimensional  lifting:\vspace{-.1cm}
\begin{subequations}
\begin{equation}\label{eq:dt_koop_general_unfact}
\Phi(x_{k+1})=A\Phi(x_k)+\mathcal{B}(x_k,u_k), \vspace{-.1cm}
\end{equation}
with $A\in\mathbb{R}^{n_{\mathrm{f}}\times n_{\mathrm{f}}}$ and\vspace{-.3cm}
\begin{multline}\label{eq:dt_B_koop_general}\vspace{-.2cm}
\mathcal{B}(x_k,u_k) = \\
\left(\int^1_0\!\frac{\partial \Phi}{\partial x}(f(x_k)\!+\!\lambda g(x_k,u_k))\dif \lambda\right)g(x_k,u_k).\vspace{-.2cm}
\end{multline}
\end{subequations}
\end{theorem_th}
\vskip -1mm
\begin{proof_th}
As $\mathbb{X}$ is convex, for any two states $p,q\in \mathbb{X}\subseteq\mathbb{R}^{n_\mathrm{x}}$ the segment $x(\lambda)=p+\lambda(q-p)$ is in $\mathbb{X}$ for all $\lambda\in[0,1]$. Next, as $\phi_i$, the $i^{\text{th}}$ component of $\Phi$ with $i\in\{1,\dots ,n_\mathrm{f}\}$,
is assumed to be continuously differentiable, define the $\mathcal{C}^1$ function $h_i:\mathbb{R}\rightarrow\mathbb{R}$ as 
$h_i(\lambda)=\phi_i \circ x(\lambda)$. 
Using the FTC (see Appendix \ref{apx:ftc}), we have 
\vspace{-.3cm}
\begin{equation}\label{eq:FTC_gi}
h_i(1) - h_i(0) = \int^1_0 h_i'(\lambda) \dif\lambda ,\vspace{-.3cm}
\end{equation}
where $h_i'=\frac{\partial h_i}{\partial \lambda}$. Next, substitute $\phi_i$ into \eqref{eq:FTC_gi} and apply the chain rule to get\vspace{-.4cm}
\begin{equation}
\phi_i(q) - \phi_i(p) = \int^1_0\frac{\partial \phi_i }{\partial x}(x(\lambda))\frac{\partial x}{\partial \lambda}(\lambda)\dif\lambda\vspace{-.2cm}
\end{equation}
which in fact provides that\vspace{-.2cm}
\begin{align}\label{eq:general_phi_i_pq}\vspace{-.2cm}
\phi_i(q) - \phi_i(p) &= \left(\int^1_0\frac{\partial \phi_i }{\partial x}(x(\lambda))\dif \lambda\right)(q - p)\\
&= \left(\int^1_0\frac{\partial \phi_i }{\partial x}(p+\lambda(q-p))\dif \lambda\right)(q - p). \notag 
\end{align}
\vskip -3mm
By choosing $q_{k+1}=f(x_k)+g(x_k,u_k)=x_{k+1}$ and $p_{k+1}=f(x_k)$ we get:\vspace{-.2cm}
\begin{align}\label{eq:DT_FTC_x_pq}\vspace{-.2cm}
x_{k+1}(\lambda)&=p_{k+1}+\lambda(q_{k+1}-p_{k+1}) \notag \\[1mm] &=f(x_k)+\lambda g(x_k,u_k).\vspace{-.2cm}
\end{align}
\vskip -2mm
Substituting \eqref{eq:DT_FTC_x_pq} into \eqref{eq:general_phi_i_pq} at time moment $k+1$ gives\vspace{-.2cm}
\begin{multline} 
\phi_i(x_{k+1}) =\phi_i(f(x_k))\ + \\
\left(\int^1_0\frac{\partial \phi_i }{\partial x}(f(x_k)+\lambda g(x_k,u_k))\dif \lambda\right)g(x_k,u_k). \vspace{-.2cm}
\end{multline}
\vskip -3mm
Stacking all components of $\Phi$ and using \eqref{eq:dt_koop_aut_gen}, the exact lifted representation of \eqref{eq:dt_nl_general} is given by 
\eqref{eq:dt_koop_general_unfact} with the input matrix function given by \eqref{eq:dt_B_koop_general}. \hfill $\blacksquare$
\end{proof_th}
In order to obtain a useful LPV 
form of \eqref{eq:dt_koop_general_unfact} as in the CT case, we can factorize  $\mathcal{B}$ in \eqref{eq:dt_koop_general_unfact}  to get 
\vspace{-.2cm}
\begin{equation}\label{eq:dt_koop_gen_fact}
\Phi(x_{k+1})=A\Phi(x_k)+B(x_k,u_k)u_k.\vspace{-.2cm}
\end{equation}
As  $g(x_k,0)=0$ by construction 
in \eqref{DT:sep}, Lemma \ref{lma_fact} gives 
\vspace{-.2cm} 
\begin{equation}
B(x_k,u_k)=\int^1_0 \frac{\partial \mathcal{B}}{\partial u}(x_k,\lambda u_k)\dif \lambda.\vspace{-.1cm}
\end{equation}
Let $z_k=\Phi(x_k)$. Similar to the CT case, the lifted form \eqref{eq:dt_koop_gen_fact} can be rewritten as the LPV form of the Koopman representation of \eqref{eq:dt_nl_general} :\vspace{-.2cm}
\begin{equation}\label{eq:lpv_dt_koop_general}
z_{k+1} = Az_k + B_{\mathrm{z}}(p_k)u_k,\vspace{-.2cm}
\end{equation}
where the scheduling map $p_k=\mu(z_k,u_k)$ is introduced such that $B_\mathrm{z} \circ \mu = B$.
\subsection{Control-affine or linear input case}
For a control-affine nonlinear system given by:\vspace{-.2cm}
\begin{equation}
x_{k+1} = f(x_k) + g(x_k)u_k,\vspace{-.2cm}
\end{equation}
Theorem \ref{thm:theorem_dt} can be applied to obtain the lifted form \eqref{eq:dt_koop_gen_fact}, where 
$B(x_k,u_k)$ simplifies as \eqref{eq:dt_B_koop_general} directly reduces to \vspace{-.3cm} 
\begin{equation}\label{eq:dt_B_koop_ca}\vspace{-.2cm}
\begin{split}
\mathcal{B}(x_k,&u_k) = \\ &\underbrace{\left(\int^1_0\frac{\partial \Phi }{\partial x}(f(x_k)+\lambda g(x_k)u_k)\dif \lambda\right)g(x_k)}_{B(x_k,u_k)}u_k.\vspace{-.2cm}
\end{split}
\end{equation}
\vskip -2mm
The input 
can easily be factorized out in this case and the LPV form of the Koopman representation follows as in
\eqref{eq:lpv_dt_koop_general}. 
For nonlinear systems with linear input: \vspace{-.2cm}
\begin{equation}
x_{k+1}=f(x_k) + bu_k, 
\end{equation}
the application of Theorem \ref{thm:theorem_dt} leads to:\vspace{-.2cm}
\begin{equation}\label{eq:dt_B_koop_lin}\vspace{-.2cm}
\mathcal{B}(x_k,u_k) = \underbrace{\left(\int^1_0\frac{\partial \Phi }{\partial x}(f(x_k)+\lambda bu_k)\dif \lambda\right)b}_{B(x_k,u_k)}u_k.\vspace{-.2cm}
\end{equation}
with an LPV form as in \eqref{eq:lpv_dt_koop_general}. 
It is important to note that, compared to the CT case, for both the control-affine and input-linear systems, the input matrix function $B$ also has a dependency on the input $u_k$ in the DT case. 
\vspace{-.1cm}
\section{Approximation error of LTI Koopman forms}\label{sec:section_error}
\smallskip
\vspace{-.1cm}
In this section we investigate how much approximation error is introduced if instead of the exact LPV Koopman representation for nonlinear systems with input, one uses only strictly LTI approximations. Such LTI Koopman forms are of interest as they are widely assumed in practice \cite{Korda:18,Mamakoukas:20,Ping:21}, however, a clear characterization of the involved approximation error is lacking in the literature. Here, we focus only on DT systems for brevity, but similar results can be obtained for the CT case. 
\subsection{Notation}
\smallskip
\vspace{-.1cm}
We introduce the following mathematical notation. $\rho(A)=\max_{r\in\lambda (A)} |r|$ denotes the spectral radius of a matrix $A\in\mathbb{R}^{n\times n}$ with eigenvalues $\lambda (A)$ and $\bar{\sigma}(P)$ is the maximum singular value of $P\in\mathbb{R}^{m\times n}$. $\|v\|_2$ is the Euclidean norm of a real vector $v\in\mathbb{R}^n$ and $\|P\|_{2,2}$ represents the induced 2,2 matrix norm:\vspace{-.3cm}
\begin{equation}\label{eq:norm_def}
\|P\|_{2,2} = \sup_{v\in\mathbb{R}^n \backslash 0}\frac{\|Pv\|_2}{\|v\|_2}=\bar{\sigma}(P).\vspace{-.4cm}
\end{equation}
For a DT signal $v:\mathbb{Z}^{+}_{0}\rightarrow \mathbb{R}^{n}$, $\|v\|_{\ell_2}=\sqrt{\sum^{\infty}_{k=0}\|v_k\|^2_2}$, where $v_k\in\mathbb{R}^n$ is the value of $v$ at time $k$, $\mathbb{Z}^{+}_{0}$ stands for non-negative integers, and $\|v\|_{\ell_\infty}=\max_{k\in\mathbb{Z}^+_0} \|v_k\|_2$.
\subsection{Characterization of the approximation error}
\smallskip
The approximate LTI Koopman model is given by:\vspace{-.3cm}
\begin{equation}\label{eq:lti_form_err}
\hat{z}_{k+1}=A\hat{z}_k + \hat{B} u_k.\vspace{-.3cm}
\end{equation}
where $\hat{z}_k$ is the associated state vector. The state matrix $A$ satisfies the embedding condition \eqref{eq:dt_koop_aut}, but $\hat{B}$ is obtained via either an approximation of $B_\mathrm{z}$ or by a data-driven scheme like \textit{(extended) dynamic mode decomposition with control} (EDMDc), see \cite{Korda:18,Proctor:16}. Let $e_k = z_k - \hat{z}_k$, where $z_k$ is the state of the exact Koopman form \eqref{eq:lpv_dt_koop_general}. Denote $B_k=B_\mathrm{z}(p_k)=B_\mathrm{z}(\mu(\Phi(x_k),u_k))$. Given the initial conditions $z_0=\hat{z}_0$ such that $e_0=0$, the error dynamics between the exact LPV Koopman form \eqref{eq:lpv_dt_koop_general} and the LTI Koopman form \eqref{eq:lti_form_err} are\vspace{-.3cm}
\begin{equation}\label{eq:err_dynamics}
e_k = Ae_{k-1} + (B_{k-1} - \hat{B})u_{k-1}.\vspace{-.3cm}
\end{equation}
In order to characterize the expected magnitude of $e_k$ as $k$ evolves, we formulate the following theorem. 
\smallskip
\begin{theorem_th}
Consider the exact Koopman embedding \eqref{eq:lpv_dt_koop_general} of a general nonlinear system \eqref{eq:dt_nl_general} and the approximative LTI Koopman form  \eqref{eq:lti_form_err}. Under any initial condition $z_0=\Phi(x_0)=\hat{z}_0$ and input trajectory $u:\mathbb{Z}^{+}_{0} \rightarrow \mathbb{R}^{n_\mathrm{u}}$ with bounded $\|u\|_{\ell_\infty}$, the state evolution error $e_k$ between these representations 
given by \eqref{eq:err_dynamics} satisfies 
\begin{enumerate}[label=(\roman*)]
\item If $\rho(A)<1$, $\|e_k\|_2$ is finite for any $k\in\mathbb{Z}^+_0$ and $\lim_{k\rightarrow\infty}\|e_k\|_2$ exists; \label{item:err_bound_i} \vspace{1mm}
\item If $\bar{\sigma}(A) < 1$, \ref{item:err_bound_i} is satisfied and furthermore\vspace{-.2cm}
\begin{equation} \label{error:bound:2}
\|e_k\|_2\leq \frac{\beta}{1-\bar{\sigma}(A)}\|u\|_{\ell_\infty},\vspace{-.2cm}
\end{equation}
where $\beta=\max_{x\in\mathbb{X},u\in\mathbb{U}} \|B_\mathrm{z}(\mu(\Phi(x),u))-\hat{B}\|_{2,2}$.
\end{enumerate}
\end{theorem_th}
\vspace{.25cm}
\begin{proof_th}
By iterative substitution, we obtain: \vspace{-.3cm}
\begin{equation}
e_k =\sum^{k-1}_{l=0}A^{k-1-l}(B_l-\hat{B})u_l.\vspace{-.3cm}
\end{equation}
Applying the 2-norm and using the triangle inequality:\vspace{-.3cm}
\begin{equation}\label{eq:inequality_err}
\|e_k\|_2 \leq \sum^{k-1}_{l=0}\|A^{k-1-l} (B_l - \hat{B})u_l\|_2.\vspace{-.3cm}
\end{equation}
Due to the submultiplicative property of \eqref{eq:norm_def}, \eqref{eq:inequality_err} can further be expanded as:\vspace{-.3cm}
\begin{equation}
\|e_k\|_2 \leq \sum^{k-1}_{l=0}\|A^{k-1-l} (B_l - \hat{B})\|_{2,2} \| u_l\|_2.\vspace{-.3cm}
\end{equation}
Using $\|u\|_{\ell_\infty}=\max_{l \in\mathbb{Z}^+_0} \|u_l\|_2$, \vspace{-.3cm}
\begin{equation}\label{eq:intermediary_ineq}
\|e_k\|_2 \leq \|u\|_{\ell_\infty} \sum^{k-1}_{l=0}\|A^{k-1-l} (B_l - \hat{B})\|_{2,2}.\vspace{-.3cm}
\end{equation}
Let $\beta=\max_{x\in\mathbb{X},u\in\mathbb{U}} \|B_\mathrm{z}(\mu(\Phi(x),u))-\hat{B}\|_{2,2}$, then the inequality \eqref{eq:intermediary_ineq} becomes:\vspace{-.4cm}
\begin{equation}\label{eq:ineq_final}
\|e_k\|_2 \leq \beta \|u\|_{\ell_\infty} \sum^{k-1}_{l=0}\|A^{k-1-l}\|_{2,2}.
\vspace{-.3cm}
\end{equation}
Under the assumption that $\rho(A)<1$, 
$\lim_{r\rightarrow \infty} A^r =0$, 
which, together with \eqref{eq:ineq_final}, show that $\|e_k\|_2$ 
is bounded for all $k\in\mathbb{Z}_0^+$, proving the first part of the theorem.

 For the second part, let $\bar{\sigma}(A)<1$. Using $\|A\|_{2,2}=\bar{\sigma}(A)$ and the closed-form of the resulting geometric series, the error can be explicitly bounded by \eqref{error:bound:2}.
\hfill $\blacksquare$
\end{proof_th}
If $\beta$ is large, which can be expected in case of significant input nonlinearities, the error \eqref{error:bound:2} can be substantial.\par
Given the LPV form of the exact Koppman representation, the computation of $\beta$ can be accomplished in a conservative manner based on a polytopic test or by gridding. Usually, $\hat{B}$ is computed via EDMDc which comes down to a least-squares approximation based on given trajectories of $x$ and $u$. However, in terms of the LPV form \eqref{eq:lpv_dt_koop_general}, one can also synthesise $\hat{B}$ by minimizing the $\ell_2$-gain of \eqref{eq:err_dynamics} 
using convex optimization \cite{IacobLPVS:22}. 
\section{Examples}\label{sec:examples}
\smallskip
In this section, to demonstrate the applicability of the introduced results, we study and compare the simulation results of various Koopman models to the original system trajectories for both CT and DT examples. Furthermore, we 
assess the tightness of the proposed error bound for LTI Koopman approximations.
\subsection{Continuous-time case}
\smallskip
In this subsection, we drop the subscript $t$ to simplify the notation. Consider the following nonlinear system:\vspace{-.3cm}
\begin{equation}\label{eq:nl_example_ct_orig}\vspace{-.2cm}
\dot{x}=\begin{bmatrix}
\mu x_1 - x_1 \\ \lambda(x_2 - x^2_1) - x_2
\end{bmatrix}+\begin{bmatrix}
x_1e^{u_1} \\ u_1u_2 + x_2e^{u_2}
\end{bmatrix},\vspace{-.2cm}
\end{equation}
where $x_i$ and $u_i$ denote the $i^{\text{th}}$ elements of the state $x$ and input $u$ vectors, respectively. For simulations, the coefficient values $\mu=-0.05$, $\lambda=-1$ are used. By computing the separation \eqref{eq:ct_general_sum}, we get
\vspace{-.2cm}
\begin{equation}\label{eq:nl_example_ct}
\dot{x}=\underbrace{\begin{bmatrix}
\mu x_1 \\ \lambda(x_2 - x^2_1)
\end{bmatrix}}_{f_\mathrm{c}(x)}+\underbrace{\begin{bmatrix}
x_1e^{u_1} - x_1 \\ u_1u_2 + x_2e^{u_2}-x_2
\end{bmatrix}}_{g_\mathrm{c}(x,u)}.\vspace{-.2cm}
\end{equation}
\vskip -2mm
The choice of observables $\Phi^\top\!(x)=[\phi_1(x)\; \phi_2(x)\; \phi_3(x)]=[x_1\; x_2\; x^2_1]$ generates a Koopman invariant subspace, such that the autonomous dynamics are represented through an exact finite-dimensional lifting\footnote{This example can be extended to a larger class of polynomial systems, for which it is possible to construct a finite dimensional
Koopman embedding irrespective of the system order (see \cite{Iacob:23} where an algorithm is given for the
embedding).}. Using \eqref{eq:ct_B_tilde_general}, the lifted representation of \eqref{eq:nl_example_ct} becomes\vspace{-.3cm}
\begin{equation}\vspace{-.2cm}
\dot{\Phi}(x)=\underbrace{\begin{bmatrix}
\mu & 0 & 0 \\ 0 & \lambda & -\lambda\\ 0 & 0 & 2\mu
\end{bmatrix}}_{A}\Phi(x)+\underbrace{\begin{bmatrix}
x_1e^{u_1} - x_1 \\ u_1u_2 + x_2e^{u_2}-x_2 \\ 2x^2_1e^{u_1}-2x^2_1
\end{bmatrix}}_{\mathcal{B}(x,u)}.\vspace{-.2cm}
\end{equation}
Next, we apply the factorization \eqref{eq:lma_fact_eq} to obtain\vspace{-.2cm}
\begin{equation}
B(x,u)=\begin{bmatrix}
\frac{x_1}{u_1}e^{u_1}-\frac{x_1}{u_1} & 0\\
\frac{1}{2}u_2 & \frac{1}{2}u_1+\frac{x_2}{u_2}e^{u_2}-\frac{x_2}{u_2}\\
2\frac{x^2_1}{u_1}e^{u_1}-2\frac{x^2_1}{u_1} & 0
\end{bmatrix}.\vspace{-.2cm}
\end{equation}
\vskip -2mm Note that when elements of $u$ become $0$ at specific time instances, $B(x,0)$ is still well defined, as it is equal to the Jacobian $\frac{\partial \mathcal{B}}{\partial u}\vert_{(x,0)}$. For example, if both $u_1=0,\;u_2= 0$, then $B^\top\!(x,0) = {\tiny \begin{bmatrix}
x_1 & 0 &2x^2_1 \\ 0 & x_2  & 0 
\end{bmatrix}}$.

The resulting lifted form of 
\eqref{eq:nl_example_ct_orig} is:\vspace{-.2cm}
\begin{subequations}\label{eq:koop_example_lifted_ct}
\begin{align} 
\dot{\Phi}(x)&=A\Phi(x) + B(x,u)u\\
x &= C\Phi(x).\vspace{-.2cm}
\end{align}
\end{subequations}
\vskip -2mm
By $z=\Phi(x)$, the LPV form of \eqref{eq:koop_example_lifted_ct} becomes\vspace{-.2cm}
\begin{subequations} \label{eq:koop_example_ct}
\begin{align} 
\dot{z}&=Az + B_\mathrm{z}(p)u\\
x &= Cz
\end{align}
\end{subequations}
\vskip -2mm
with $p =\mu(z,u)= [z^\top \; u^\top]^\top$, $B_\mathrm{z} \circ \mu=B$ and $C=[I_2 \; 0_{2\times 1}]$. As $B$ contains functions with only linear dependencies on the observables $\Phi(x)$, computing $B_\mathrm{z}$ is trivial and is omitted here for brevity. 

\begin{figure}[t!]
\begin{center}
	\includegraphics[width=.45\textwidth,trim={.15cm .25cm .3cm .25cm},clip]{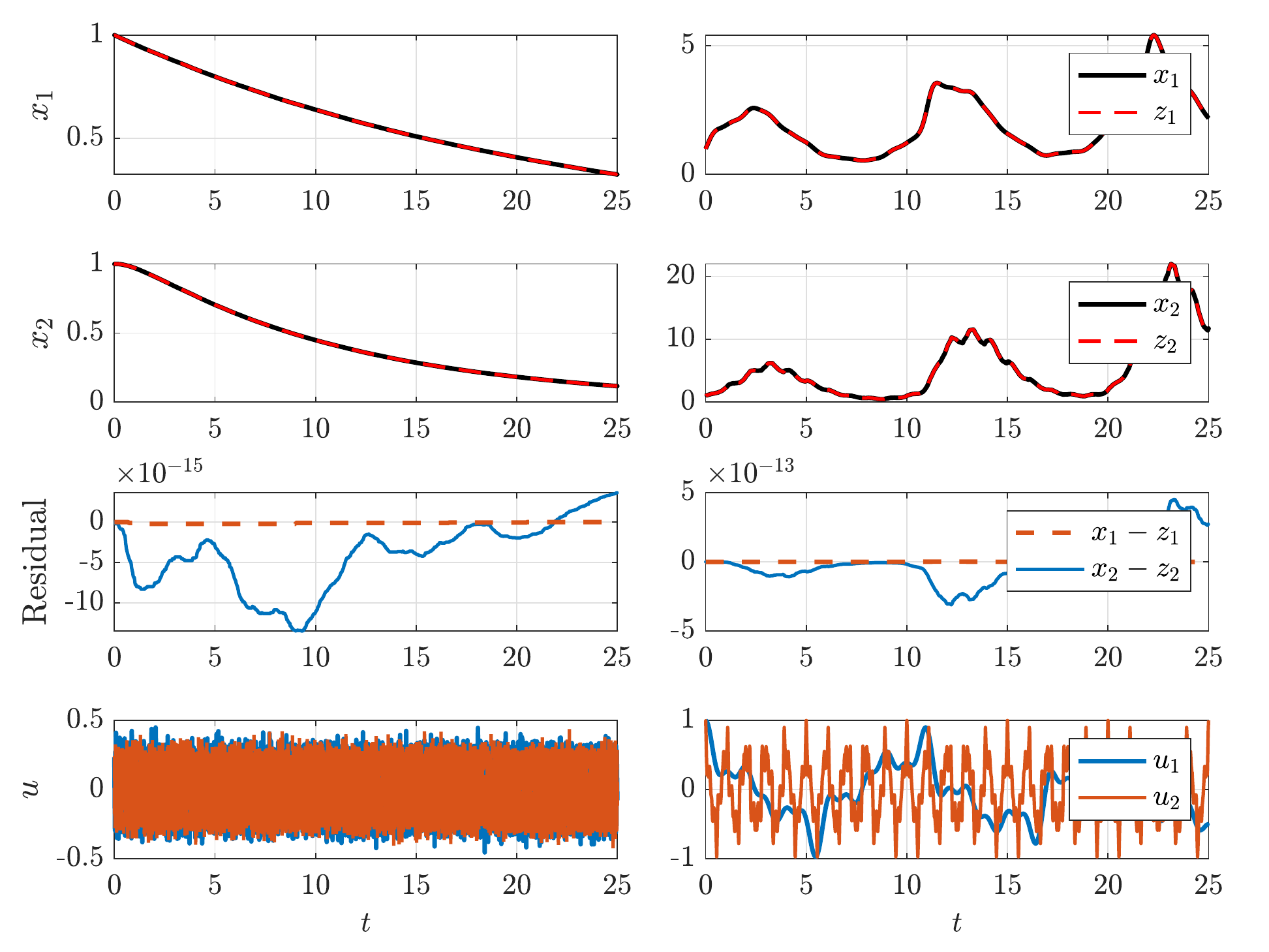}    
\vspace{-3mm}\caption{Continuous-time example: state response of the original nonlinear system \eqref{eq:nl_example_ct_orig} given in black  and its exact Koopman representation \eqref{eq:koop_example_ct} given in red under white noise (left panel) and multisine (right panel) excitation $u$.}\vspace{-.1cm} 
\label{fig:ct_fact_example}
\end{center}
\end{figure}

To compare the responses of the original system and the Koopman form, the ODEs \eqref{eq:nl_example_ct_orig} and \eqref{eq:koop_example_ct} are solved by \emph{Runge-Kutta 4} (RK4) with a sampling time $T_\mathrm{s}=10^{-4}$s  under initial condition $x_0=[1\;1]^\top$, corresponding to $z_0=\Phi(x_0)=[1\;1\;1]^\top$, and white input signals $u_i(t_k) \sim \mathcal{N}(0,0.1)$ (where $t_k$ represents the discrete time step of the numerical integration). The same simulations have been repeated under multisine inputs (no phase difference) with 6 excited frequencies equidistantly placed on the frequency range $[0.1, 1]$ Hz for $u_1$, and $[1,10]$ Hz for $u_2$, respectively. To quantify the approximation quality of the model we use the $\|\cdot\|_{\ell_2}$ and $\|\cdot\|_{\ell_\infty}$ norms of the difference between the $i^{\text{th}}$ component of the state evolution, i.e., $\epsilon_i(t_k)=x_i(t_k)-z_i(t_k)$. We collect in $\epsilon_i$ the differences for all times, i.e., $\epsilon_i=[\epsilon_i(t_0)\ \cdots\  \epsilon_i(t_N)]$, with $i\in \left\lbrace 1,2 \right\rbrace$.
 Fig. \ref{fig:ct_fact_example} shows the state trajectories for the two simulations for a time length of 25 seconds. Visually, there is a perfect overlap between the state trajectories computed via the original nonlinear representation \eqref{eq:nl_example_ct_orig} and the Koopman representation \eqref{eq:koop_example_ct}. As detailed in Table \ref{tab:Error_measures}, under both white noise and multisine excitation, the error measures $\|\epsilon_i\|_{\ell_2}$ are below $10^{-10}$ and $\|\epsilon_i\|_{\ell_\infty}$ are below $10^{-12}$, respectively, being close to numerical accuracy. This shows that the Koopman form accurately represents the original system up to numerical precision of the simulation.

\subsection{Discrete-time case}\label{sec:dt_example}
\smallskip
In this subsection, the 
$x_{k,i}$ represents the value of the $i^{\text{th}}$ state at time $k$. Consider the nonlinear system represented by the following  control-affine state-space form\vspace{-.2cm}
\begin{equation}\label{eq:nl_dt_ex}
x_{k+1} = \underbrace{\begin{bmatrix}
a_1x_{k,1} \\ a_2x_{k,2} - a_3x^2_{k,1}
\end{bmatrix}}_{f(x_k)}+ \underbrace{\begin{bmatrix}
1 \\ x^2_{k,1}
\end{bmatrix}}_{g(x_k)}u_k. \vspace{-.2cm}
\end{equation}
\vskip -2mm
For simulation purposes, the used coefficient values are $a_1=a_2=0.7$ and $a_3=0.5$. Using the same observable choice as for the continuous-time example $\Phi^\top (x_k)=[\phi_1(x_k)\; \phi_2(x_k)\; \phi_3(x_k)]=[x_{k,1}\; x_{k,2}\; x^2_{k,1}]$ yields an exact finite-dimensional lifting of the autonomous part. 
Based on \eqref{eq:dt_koop_gen_fact} with \eqref{eq:dt_B_koop_ca}, the lifted form is: \vspace{-.2cm}
\begin{subequations}
\label{eq:lifted_model_example_dt}
\begin{align}
\Phi(x_{k+1})&\!=\!\!\underbrace{\scriptsize \begin{bmatrix}
a_1 & 0 & 0 \\ 0 & a_2 & -a_3 \\ 0 & 0 & a^2_1
\end{bmatrix}}_{A}\!
\Phi(x_k)\!+\!\!\underbrace{\scriptsize \begin{bmatrix}
1 \\ x^2_{k,1} \\ 2a_1x_{k,1}+u_k \end{bmatrix}}_{B(x_k,u_k)}\!u_k,\\
x_k&=C\Phi(x_k).
\end{align} 
\end{subequations}
\begin{table*}[ht!]
\parbox{16.5cm}{\caption{Characterization of the state-evolution error between the original nonlinear system and the Koopman forms in the considered simulation examples. \label{tab:Error_measures}}} \vspace{.1cm}
\begin{center}
 \begin{tabular}{||c | c | c | c | c | c | c | c ||} 
 \hline
  \multirow{2}{*}{Input} & \multirow{2}{*}{$x_i$} & \multicolumn{2}{|c|}{Continuous time}  & \multicolumn{2}{|c}{Discrete time}  & \multicolumn{2}{|c||}{Discrete time with constant $\hat{B}$} \\ \cline{3-8} & & $\|\epsilon_i\|_{\ell_2}$  & $\|\epsilon_i\|_{\ell_\infty}$ & $\|\epsilon_i\|_{\ell_2}$ &$\|\epsilon_i\|_{\ell_\infty}$ & $\|\epsilon_i\|_{\ell_2}$ & $\|\epsilon_i\|_{\ell_\infty}$ \\  
 \hline\hline
 \multirow{2}{*}{Multisine} & $i=1$ & $3.12\cdot 10^{-13}$& $1.77\cdot 10^{-15}$ &  $0$& $0$ &  $1.60\cdot 10^{-14}$ & $8.88\cdot 10^{-16}$  \\
 \cline{2-8}
  & $i=2$ & $7.71\cdot 10^{-11}$ &$4.51\cdot 10^{-13}$ &  $3.50\cdot 10^{-14}$ &$7.10\cdot 10^{-15}$ & $190.82$ &$18.06$ \\
 \hline
\multirow{2}{*}{White noise} & $i=1$ & $6.96\cdot 10^{-14}$& $2.22\cdot 10^{-16}$ &  $0$& $0$ &  $7.31\cdot 10^{-15}$ & $8.88\cdot 10^{-16}$  \\
 \cline{2-8}
  & $i=2$ & $2.96\cdot 10^{-12}$ &$1.34\cdot 10^{-14}$ &  $1.23\cdot 10^{-14}$ &$3.55\cdot 10^{-15}$ & $58.12$ &$9.17$ \\
 \hline
\end{tabular}
\end{center}
\end{table*}
Let $z_k=\Phi(x_k)$ and notice that $x^2_{k,1}$ and $x_{k,1}$ are part of the observables. Thus, we 
get $B_\mathrm{z} \circ \mu = B$ and write the Koopman representation of \eqref{eq:nl_dt_ex} in an LPV form \vspace{-.2cm}
\begin{equation}\label{eq:koop_model_example}
\begin{split}
z_{k+1}&=Az_k+B_\mathrm{z}(p_k)u_k,\\
x_k&=Cz_k,
\end{split}
\end{equation} \vskip -4mm
with $p_k = \mu(z_k,u_k)= [z_k^\top\; u_k]^\top$ and $C=[I_2 \; 0_{2\times 1}]$. The initial conditions are taken as $x_0=[1\;1]^\top$, $z_0=[1\;1\;1]^\top$.
\begin{figure}[t]
\begin{center}
\includegraphics[width=.45\textwidth,trim = {.15cm .25cm .3cm .25cm}, clip]{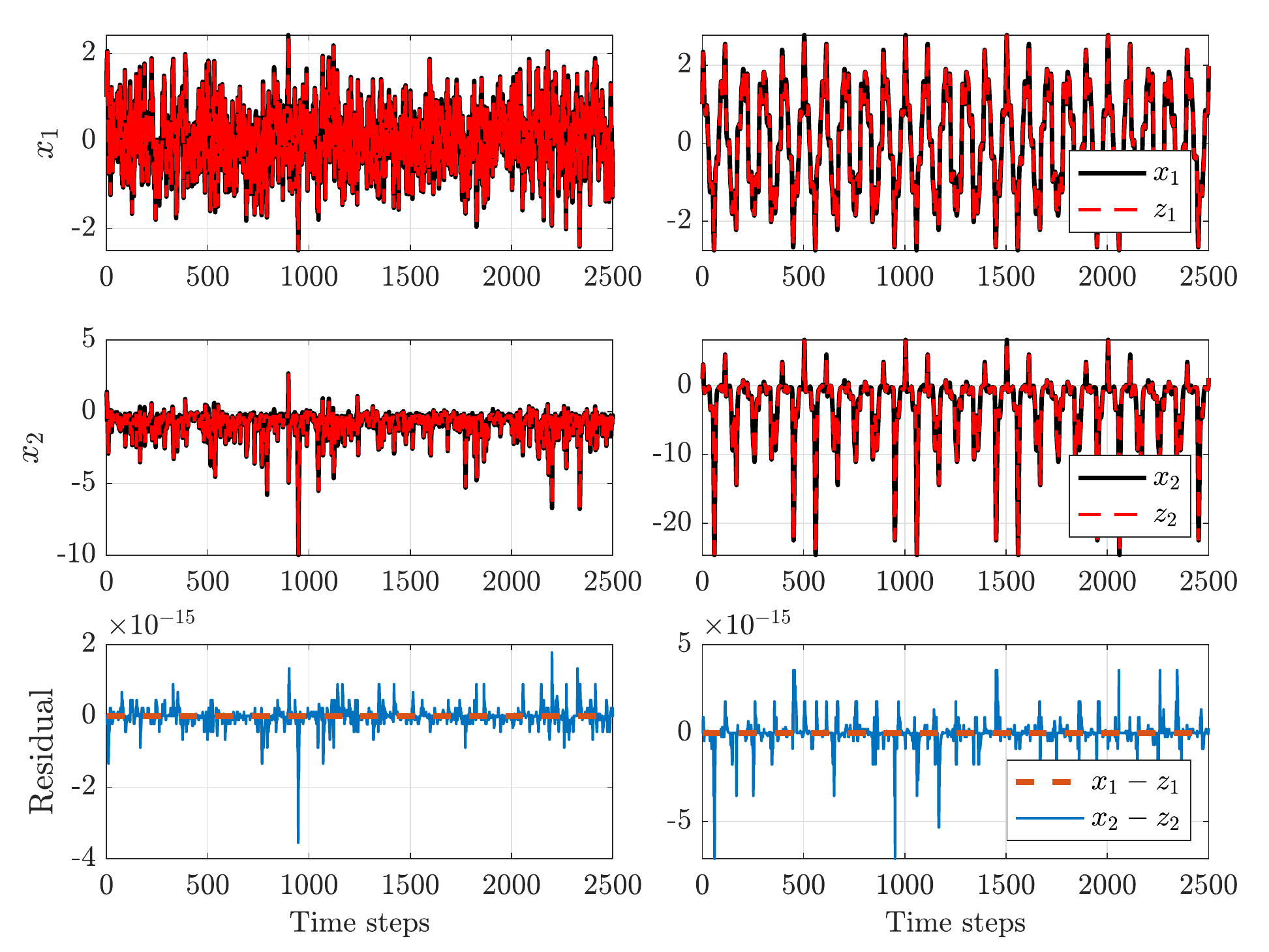}  
\vspace{-3mm}
\caption{Discrete-time example: state response of the original nonlinear system \eqref{eq:nl_dt_ex} given in black  and its exact Koopman representation \eqref{eq:koop_model_example} given in red under white noise (left panel) and multisine (right panel) excitation $u$.} \vspace{-.1cm} 
\label{fig:dt_ca_example}
\end{center}
\end{figure}
Fig. \ref{fig:dt_ca_example} shows the state trajectories computed via the nonlinear model \eqref{eq:nl_dt_ex} and the Koopman representation \eqref{eq:koop_model_example}, for white noise $u_k\sim \mathcal{N} (0, 0.5)$ and for multisine excitation as well. Similar to the CT case, the multisine input is a sum of 6 sinusoids (no phase difference) with a constant increment between the excited frequencies. In a similar manner as in the previous example, we denote by $\epsilon_i$ the vector containing the differences between the $i^{\text{th}}$ element of the state evolution, i.e., $\epsilon_i=[\epsilon_{0,i}\ \cdots\ \epsilon_{N,i}]$ and $\epsilon_{k,i}=x_{k,i}-z_{k,i}$, with $i\in \left\lbrace 1,2 \right\rbrace$. As shown in Table \ref{tab:Error_measures}, the first state is represented exactly and, for the second state, there are negligible deviations which are close to numerical precision. This shows that the general Koopman representation accurately describes \eqref{eq:nl_dt_ex}. \vspace{-.1cm}
\subsection{Approximation by an LTI Koopman form}
\smallskip
\vspace{-.1cm}
While it has been shown that the proposed approach yields an exact  finite-dimensional Koopman form that accurately represents the dynamics of the original system, the input matrix is dependent on the state (and on the input, in the DT case). One can wonder if such dependency really contributes to the system response and if one could get away with a fully LTI Koopman form as it is done in the works \cite{Korda:18,Mamakoukas:20,Ping:21}. As we show in this section, the accuracy of the resulting model can drastically decrease if only an LTI approximation is used.
\begin{figure}[!t]
\vspace{-0.2cm}
\begin{center}
\includegraphics[width=.45\textwidth,trim = {.15cm .25cm .3cm .25cm}, clip]{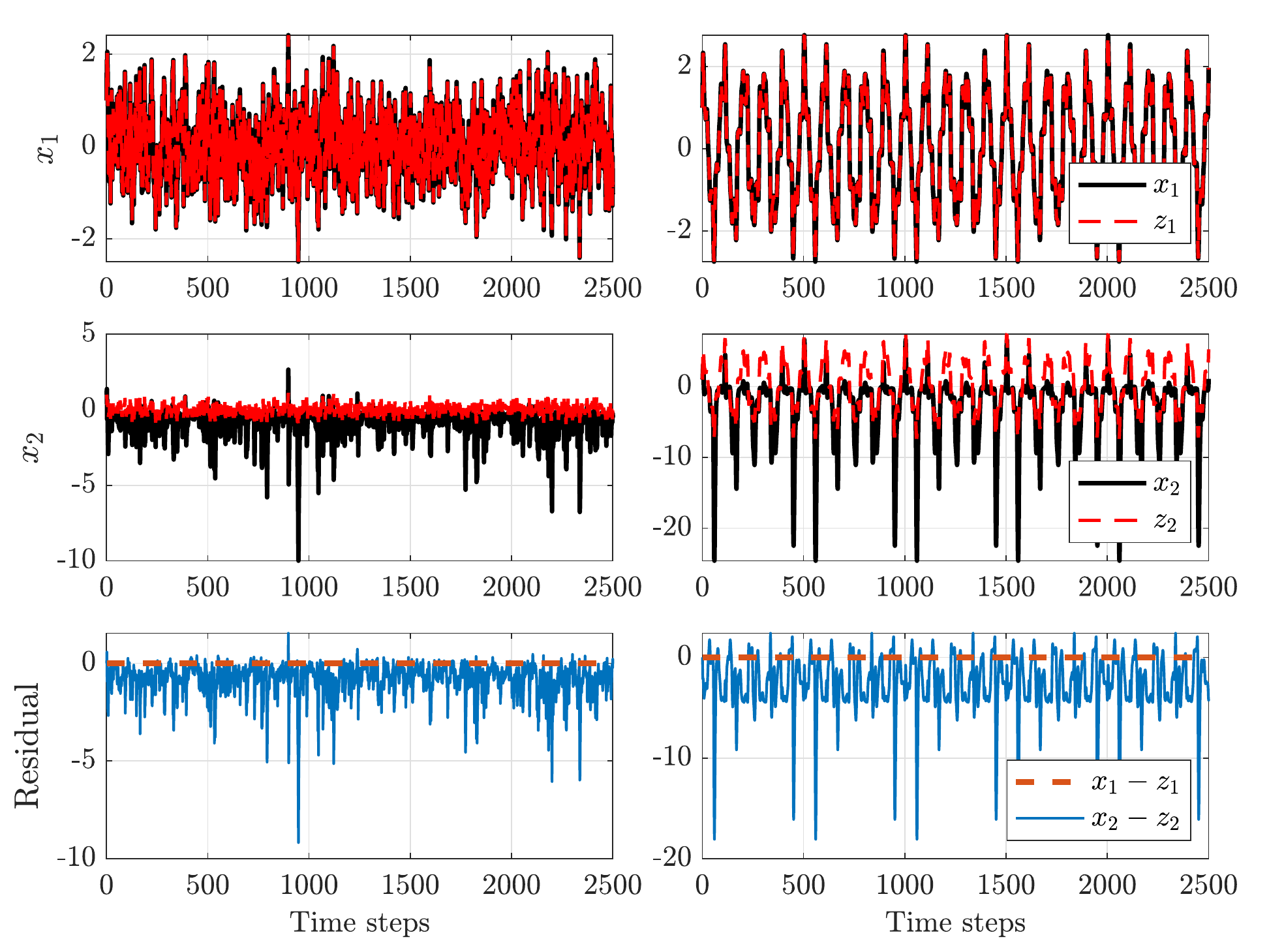}\vspace{-.3cm}    
\caption{Comparison to the approx. LTI form: state response of the nonlinear system \eqref{eq:nl_dt_ex} given in black  and its approx. LTI Koopman model \eqref{eq:koop_fixed_B} given in red under white noise (left panel) and multisine (right panel) excitation $u$.}\vspace{-.2cm} 
\label{fig:dt_side_fixed_B}
\end{center}
\end{figure}
\par
\begin{figure}[t]
\vspace{-.25cm}
\begin{center}
\includegraphics[width=.45\textwidth,trim = {.15cm .25cm .3cm .25cm}, clip]{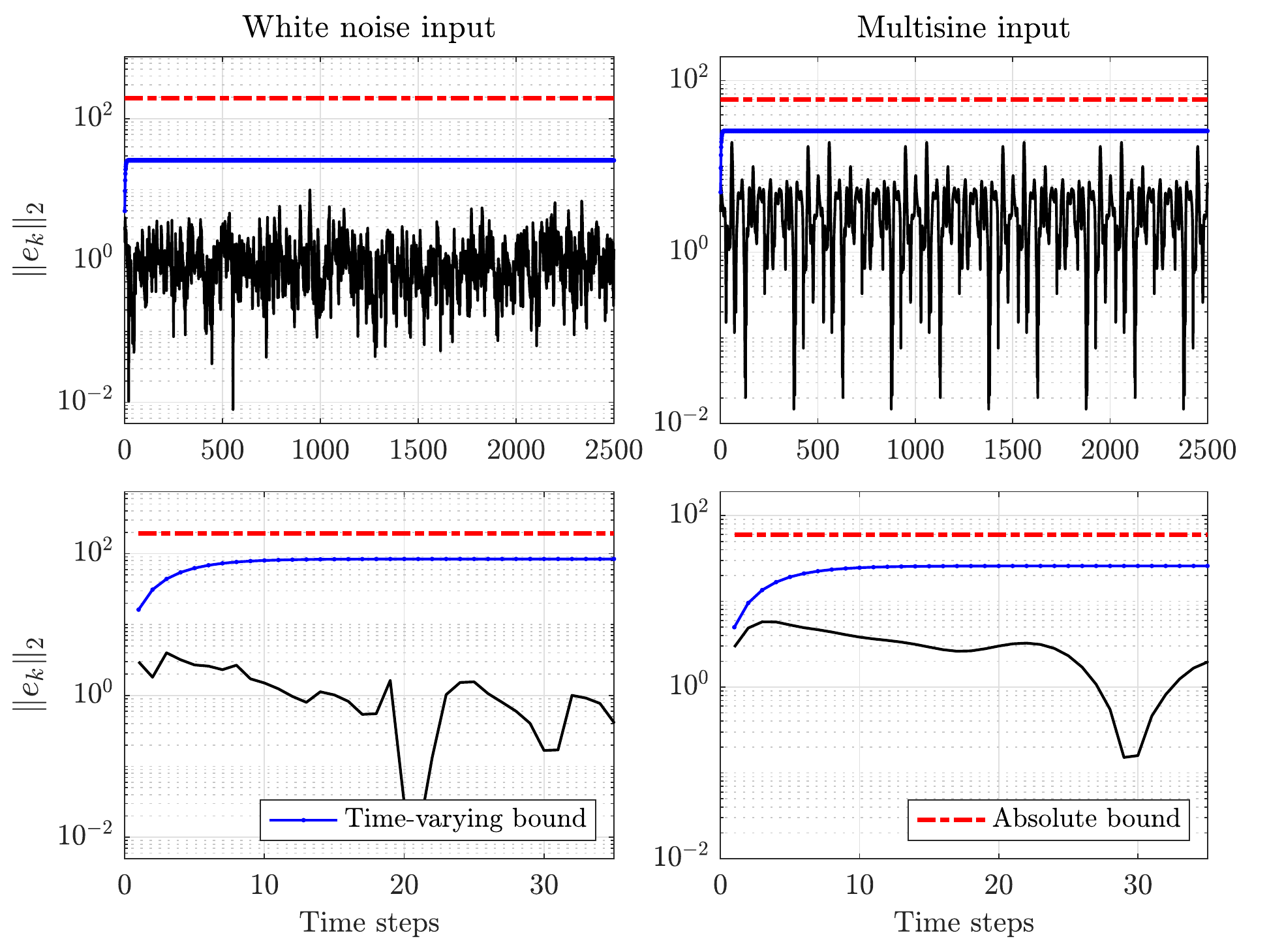}    
\vspace{-3mm} \caption{Evolution of the state-response error $\|e_k\|_2$, the time-varying error bound \eqref{eq:ineq_final} (blue), and the absolute error bound \eqref{error:bound:2} (red) for the approximative LTI Koopman model under white noise (left panel) and multisine inputs (right panel). Zoomed in view of the initial time steps is provided in the second row.} \vspace{-.2cm}
\label{fig:error_bound}
\end{center}
\end{figure}

 Consider the DT example \eqref{eq:nl_dt_ex}. With a constant input matrix $\hat{B}$, the assumed Koopman form of the system is:\vspace{-.2cm} 
\begin{equation}\label{eq:koop_fixed_B}
z_{k+1} = Az_k + \hat{B}u_k.\vspace{-.2cm}
\end{equation}
The original states, being part of the observables, are recovered through $x_k=[I_2\; 0_{2\times 1}]z_k$. One typical way to find $\hat{B}$ is to minimize the average 2-norm between $B$ and $\hat{B}$ by the following approach. To get the most favorable computation of $\hat{B}$ for comparison, we take the grid points in $(\mathbb{X},\mathbb{U}$) corresponding to a simulation trajectory of \eqref{eq:nl_dt_ex} and formulate the matrices $Z=[
\ \Phi(x_0) \ \cdots \ \Phi(x_{N-1})\ ]
$, $Z^+=[ \
\Phi(x_1) \ \cdots \ \Phi(x_{N}) \ ]
$ and $U=[ \
u_0 \ \cdots \ u_{N-1} \ ]$. With $A$ derived analytically in \eqref{eq:lifted_model_example_dt}, the input matrix $\hat{B}$ is numerically computed as:\vspace{-.2cm}
\begin{equation}\label{eq:koop_edmd_exact_lift}
\hat{B}=(Z^+ - AZ)U^\dagger ,\vspace{-.2cm}
\end{equation}
where $\dagger$ denotes the pseudoinverse. This approach corresponds to the least-squares method used in EDMD \cite{Korda:18} with the difference that we only need to compute the input matrix $\hat{B}$. The simulations are repeated according to Section \ref{sec:dt_example}. Fig. \ref{fig:dt_side_fixed_B} shows large deviations between the second state trajectory, computed via the nonlinear dynamics \eqref{eq:nl_dt_ex} (black) and using the Koopman form \eqref{eq:koop_fixed_B} (red). The input term is nonlinear for the second state in \eqref{eq:nl_dt_ex}, while the first state is affected linearly by $u$ and its evolution is independent of $x_2$. Hence, the overall approximation error is expected to be larger for the second state evolution, whereas the first state is only slightly affected. The same conclusion is supported by the error measures reported in Table \ref{tab:Error_measures}. Furthermore, the evolution of the error $\|e_k\|_2$ of the approximative LTI Koopman model \eqref{eq:koop_fixed_B} is displayed in Fig. \ref{fig:error_bound} together with the error bounds \eqref{error:bound:2} and \eqref{eq:ineq_final}. It can be observed that $\|e_k\|_2$ satisfies both bounds and the time-varying bound \eqref{eq:ineq_final} converges to a fixed value that is less conservative than the absolute bound \eqref{error:bound:2}. 
\begin{figure}[t]
\begin{center}
\includegraphics[width=.45\textwidth,trim = {.15cm .05cm .3cm .05cm}, clip]{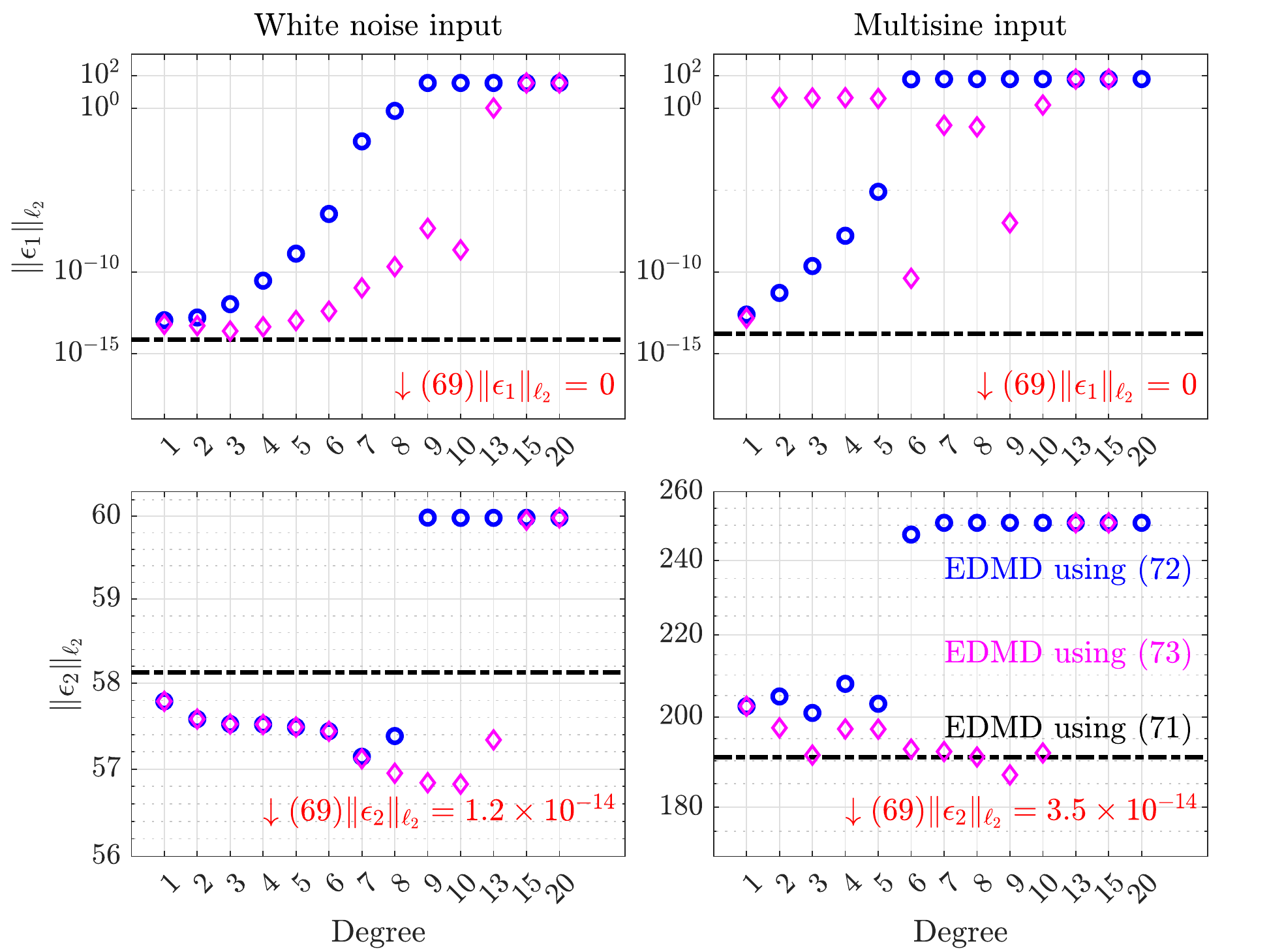}\vspace{-0.3cm}    
\caption{$\ell_2$ error of simulated state trajectories of the LTI Koopman model obtained with observables of a full dictionary of monomials with increasing degrees (blue circles), the model obtained using regularized EDMD (magenta diamonds), the model obtained via \eqref{eq:koop_edmd_exact_lift} (black line), and the exact Koopman representation \eqref{eq:koop_model_example} (red).
}\vspace{-.2cm}
\label{fig:errors_edmd}
\end{center}
\end{figure}
\par
To analyze how a large dictionary influences the  approximation error of the LTI Koopman model, we have also applied the standard EDMD with a full dictionary of monomial sets of certain degrees (terms of the form $x_1^ax_2^b$ where $a+b \leq \text{degree}$) as observables. Note that the number of distinct functions depends on the degree (e.g. for degree 3, the lifting set has 9 functions). To compute the $A$ and $B$ matrices numerically, we have employed the method detailed in \cite{Korda:18}, for larger datasets ($N \gg n_{\mathrm{f}}$),  \vspace{-.3cm}
\begin{equation}
[A\; B] = VG^{\dagger}, \; V=\begin{bmatrix}
Z\\ U
\end{bmatrix}\begin{bmatrix}
Z\\ U
\end{bmatrix}^\top\!\!\!\!, \; G=Z^+\begin{bmatrix}
Z\\ U
\end{bmatrix}^\top\!\!\!\!.\vspace{-.2cm}
\end{equation}
Note that EDMD is inherently a one step predictor and,
while increasing the dictionary size may improve results, it can also be a cause of numerical errors. Furthermore, as described in \cite{Brunton:22}, increasing the
dictionary may create spurious eigenvalues that can produce worse results in simulation based testing of the
resulting model. To avoid such problems, we have also added a Thikonov regularization term to perform the lifting.
We used the following estimator:\vspace{-.3cm}
\begin{equation}
    [A\; B] = Z^+ Y^\top (YY^\top + \alpha I)^{-1}\vspace{-.2cm}
\end{equation}
with $Y = [Z^\top\; U^\top]^\top$ and $\alpha$ the scalar regularization term. To optimize $\alpha$, a grid search is performed over a logarithmic grid scale between $[10^{-15},10^{20}]\cup \{0\}$, to minimize the cost $\|\epsilon_1\|_{\ell_2}+\|\epsilon_2\|_{\ell_2}$. The simulations are performed using the same initial values and inputs detailed in Section  \ref{sec:dt_example}. As can be observed in Fig. \ref{fig:errors_edmd}, increasing the number of observables does not bring big improvements in the approximation quality of the second state and has an adverse effect for the first state. The regularized approach improves the quality of approximation in the white noise input case, but mixed results can be observed when the system is excited with a multisine input. It can be seen that the model still fails to capture the dynamics of the original system. Also, note that for the monomial set of degree 20, the system becomes unstable in the case of the multisine input and this was omitted from the figure to increase readability. While the model with the exactly lifted autonomous part generally shows better accuracy, these results support the conclusion that LTI models can not capture the dynamics of \eqref{eq:nl_dt_ex} correctly, whereas the LPV form of the Koopman description \eqref{eq:koop_model_example} is an exact representation of the dynamics (up to machine precision).\vspace{-.1cm}
\section{Conclusion}\label{sec:conclusion}
\smallskip
\vspace{-.15cm}
We have developed a systematic approach to analytically derive a Koopman representation of both continuous and discrete-time general nonlinear systems with inputs. Furthermore, we have shown that the resulting lifted forms can be interpreted as LPV models, allowing for powerful LPV tools to be used for analysis and control of nonlinear systems. As seen through the examples, this approach results in an exact representation of the original dynamics in contrast to the often assumed purely LTI form of Koopman models which are heavily limited in their representation capability. To characterize the approximation capability of LTI Koopman models, an error bound has been derived. We have shown that in case of systems with inputs, although LTI Koopman models are inaccurate compared to the exact LPV type of Koopman representations, their simulation error remains bounded and has a predictable behaviour.
\vspace{-.2cm}\appendix
\section{Proof of the Factorization Lemma}\label{apx:proof_lemma} \smallskip \vspace{-.15cm}   
This section details the proof of Lemma \ref{lma_fact}. \smallskip
\begin{proof_th}
Let $\lambda\in\mathbb{R}$, $p,q\in \mathbb{U}\subseteq\mathbb{R}^{\mathrm{n}_u}$ and define an arbitrary input function $u$ as the convex combination $u(\lambda)=p+\lambda(q-p)$, with $u\in [p,q]$ (element wise) and $\lambda \in [0,1]$. Define the function $\zeta_i(\lambda)=\mathcal{B}_i(x,u(\lambda))$, where $\mathcal{B}_i$ denotes the $i^{\text{th}}$ row of $\mathcal{B}$ and $i\in \{1,\dots,n_\mathrm{f}\}$. By applying the FTC (see Appendix \ref{apx:ftc}), the following statement holds true:\vspace{-.4cm}
\begin{equation}
\zeta_i(1)-\zeta_i(0)=\int^1_0 \zeta_i' (\lambda)\dif \lambda,\vspace{-.45cm}
\end{equation}
where $\zeta_i'=\frac{\partial \zeta_i}{\partial \lambda}$. This is equivalent to:\vspace{-.35cm}
\begin{equation}
\mathcal{B}_i(x,q)-\mathcal{B}_i(x,p) = \int^1_0 \frac{\partial \mathcal{B}_i}{\partial u}(x,u(\lambda))(q-p) \dif \lambda.\vspace{-.35cm}
\end{equation}
Next, the row elements can be vertically stacked giving:\vspace{-.3cm}
\begin{equation}
\mathcal{B}(x,q)\!-\!\mathcal{B}(x,p) \!=\!\! \left(\int^1_0 \frac{\partial \mathcal{B}}{\partial u}(x,u(\lambda)) \dif \lambda\right)(q\!-\!p). \vspace{-.3cm}
\end{equation}
By choosing $q=u$ and $p=0$, the function $\mathcal{B}(x,0)=0$ (as $g_c(x,0)=0$) and the factorized formulation of $\mathcal{B}$ is 
$\mathcal{B}(x,u)=B(x,u)u$ 
with\vspace{-.45cm}
\begin{equation}\label{eq:ct_factorization}
B(x,u)=\int^1_0\frac{\partial \mathcal{B}}{\partial u}(x,\lambda u)\dif \lambda . \vspace{-.1cm}
\end{equation}
\end{proof_th}
\vspace{-.5cm}
\section{Fundamental Theorem of Calculus}\label{apx:ftc}
\smallskip
\vspace{-.2cm}
The following theorem and its proof can be found in \cite{Thomas:05}. We only reproduce the theorem here for completeness.
\smallskip
\begin{theorem_th}
If a function $f$ is continuous over an interval $[a,b]$ and $F$ is any antiderivative of $f$ on $[a,b]$, then:\vspace{-.4cm}
\begin{equation}
\int^b_af(x)\dif x = F(b)-F(a).\vspace{-.45cm}
\end{equation}
\end{theorem_th}
\bibliographystyle{plain}        

\bibliography{references_koopman_input_paper}         
\vspace*{1em}
\renewcommand{\baselinestretch}{0.83}
\selectfont
\vspace{-0.55cm}
\biographytwo{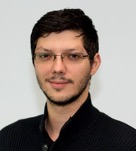}{Lucian Cristian Iacob}{received his BSc degree in Systems Engineering at the Technical University of Cluj-Napoca, in 2017. He obtained his MSc degree in Systems and Control at Eindhoven University of Technology (TU/e) in 2019, with distinction (great appreciation). He is currently pursuing a Ph.D. degree in the Control Systems Group at TU/e. His research interests are on modelling and identification of nonlinear systems for control using the Koopman and linear parameter-varying (LPV) frameworks.} \vspace{-0.35cm}
\biographytwo{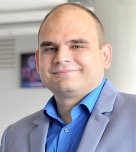}{Roland T\'oth}{received his Ph.D. degree with cum laude distinction at the Delft Center for Systems and Control (DCSC), Delft University of Technology (TUDelft), Delft, The Netherlands in 2008.  He was a Post-Doctoral Research Fellow at TUDelft in 2009 and Berkeley in 2010. He held a position at DCSC, TUDelft in 2011-12. Currently, he is an Associate Professor at the Control Systems Group, Eindhoven University of Technology and a Senior Researcher at SZTAKI, Budapest, Hungary. His research interests are in identification and control of linear parameter-varying (LPV) and nonlinear systems, developing machine learning methods with performance and stability guarantees for modelling and control, model predictive control and behavioral system theory. }\vspace{-0.35cm}
\biographytwo{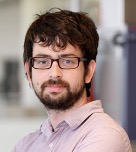}{Maarten Schoukens}{is an Assistant Professor in the Control Systems group of the Department of Electrical Engineering at the Eindhoven University of Technology (TU/e). He received his Ph.D. degree in engineering from the Vrije Universiteit Brussel (VUB), Brussels, Belgium in 2015. He has been a Post-Doctoral Researcher with the ELEC Department, VUB, and the Control Systems research group, TU/e, Eindhoven, The Netherlands. In 2018 he became an Assistant Professor in the Control Systems group, TU/e. His main research interests include the measurement and data-driven modelling and control of nonlinear dynamical systems using system identification and machine learning techniques.}
\end{document}